# Electron hole – phonon interaction and structural changes in $La_{0.9}Sr_{0.1}FeO_3$ by temperature dependent conductivity and valence band photoemission spectroscopy

A. Braun[1][*], J. Richter[1], A. S. Harvey[2], A. Infortuna[2], A. Frei[3], E. Pomjakushina[4], Bongjin S. Mun[5,6],

P. Holtappels[1], U. Vogt[1], K. Conder[4], L. J. Gauckler[2], T. Graule[1]

[1]*Laboratory for High Performance Ceramics*

*EMPA – Swiss Federal Laboratories for Materials Testing & Research*

*CH – 8600 Dübendorf, Switzerland*

[2]*Department for Non-Metallic Materials*

*ETH Zürich – Swiss Federal Institute of Technology*

*CH – 8037 Zürich, Switzerland*

[3]*Department for General Energy Research, Paul Scherrer Institut*

*CH-5232 Villigen PSI, Switzerland*

[4]*Laboratory for Development and Methods, Paul Scherrer Institut*

*CH-5232 Villigen PSI, Switzerland*

[5]*Department of Applied Physics, Hanyang University*

*Ansan, Kyunggi-Do 426-791, Korea*

[6]*Advanced Light Source, Lawrence Berkeley National Laboratory*

*Berkeley CA 94720, Berkeley, USA*

---

[1] *Corresponding author. EMPA – Swiss Federal Laboratories for Materials Testing & Research, Überlandstrasse 129, CH – 8600 Dübendorf, Switzerland, Phone +41 (0)44 823 4850, Fax +41 (0)44 823 4150, email: artur.braun@alumni.ethz.ch

**Abstract**

Electric conductivity and structural details of the hole-doped polaron conductor $La_{0.9}Sr_{0.1}FeO_3$ (LSF10) are reported. The conductivity of a single crystal shows an exponential increase with temperature with a maximum of 100 S/cm at 700 K and with activation energy of about 375 meV, and a decrease for higher temperatures that follows a power law. The exponential increase of the electric conductivity for 300 K ≤ T ≤ 700 is accompanied by a shift of spectral weight in the photoemission valence band towards the Fermi level, indicative of a strong electron-phonon interaction. The subsequent decrease of the conductivity for T > 700 K is accompanied by a reversible phase transformation from orthorhombic to rhombohedral symmetry. The decreasing conductivity for T > 700 K is likely due to the reduction of the iron due to oxygen loss causing a decreasing hole concentration, as evidenced by a substantial chemical shift in the Fe K-shell x-ray absorption spectra. Two additional fine structures in the conductivity data at 357 K, this is, a small temperature reversible jump in the conductivity, and at 573 K, a slight reversible increase of the polaron activation energy, are correlated with an exceptionally strong decrease in spectral valence band intensity near the Fermi level, and with the onset of a corresponding structural transition.

**Introduction**

Rare earth – transition metal oxides have been of interest for solid state scientists for a long time, not least for their interesting transport properties such as metal-insulator transitions, high temperature superconductivity, and colossal magnetoresistance, for example. The studies on these phenomena are typically restricted to ambient temperatures and, in case of superconductivity, significantly below 300 K. Highly conductive materials at high temperatures from 300 K to 1200 K are desirable for solid oxide fuel cells, electrolysers, oxygen separation membranes and sensors. The description of such systems at high temperatures is particularly difficult because the materials may, in addition to electron – phonon interaction, undergo redox reaction, adjusting their oxygen content to comply with the phase diagram and other thermodynamic constraints.

Lanthanum-strontium ferrites (LSF) have a distorted perovskite structure. $LaFeO_3$ has orthorhombic symmetry with four distorted pseudo-cubic cells in the space group Pbnm, and is an antiferromagnetic charge transfer insulator with 2 eV band gap energy. Each Fe-atom has its spin antiparallel to its six closest neighbouring Fe atoms. The Néel temperature is 750 K and drops when Sr is added. Parasitic ferro- and ferrimagnetism is found, including permanent magnetization [1,2], when the level of Sr doping is kept small. The *heterovalent* substitution of $La^{3+}$ in the insulating $LaFeO_3$ with $Sr^{2+}$ (hole doping) creates electron hole states with substantial O (2p) character near the Fermi level [3]. The electric conductivity of $La_{(1-x)}Sr_xFeO_3$ increases with increasing substitution parameter $x$ to the extent that $SrFeO_3$ shows typical metal behaviour. $SrFeO_3$ is an antiferromagnetic metallic conductor with temperature dependent conductivity [4].

During the hole doping process, La and Sr are considered as retaining their formal valency of $La^{3+}$ and $Sr^{2+}$. The charge balance is primarily maintained by adjustment of the oxidation state of the Fe, and secondarily by an onset of oxygen deficiency $\delta$ in the $La_{(1-x)}Sr_xFeO_3$. The formal valency of Fe in $LaFeO_3$ and $SrFeO_3$ is $Fe^{3+}$ and $Fe^{4+}$, respectively. The $Fe^{4+}$ ions represent the electron holes and control the conductivity, whereas the $Fe^{3+}$ ions control via their spin structure the magnetic properties [5]. Valence band photoemission spectroscopy (VB PES) on LSF has been carried out recently by several research groups. In this study we report on $La_{0.9}Sr_{0.1}FeO_3$ (LSF10) single crystal material, which comes close in properties to the insulating parent compound $LaFeO_3$.

**Experimental**

$La_{0.9}Sr_{0.1}FeO_3$ powder was synthesised by solid state thermal reaction. $La(OH)_3$, $SrCO_3$ and $Fe_2O_3$ were mixed in the stoichiometric amounts, heated up to 1473 K with 5 K/min and hold for 10 h at temperature, followed by a cooling rate of 5 K/min. Phase purity was confirmed by x-ray diffraction. General data on the crystallographic and electronic structure of LSF are available in [1,6,7].

To obtain a single crystal, the powder was first pressed in a cold isostatic press (Vitec) at 2000 bar to a rod of 8 mm in diameter and presintered at 1573 K for 4 h (Naber Therm HT 08/17). The seed and feed rods were pressed from the powder at 4 kbar in a hydrostatic press and sintered at 1523 K (12h) in air. The single crystal was grown in an optical floating zone furnace (CSC Japan), powered with 1000 Watt halogen tubes at a growth rate of 1.5 mm/h in argon plus 1% $O_2$ atmosphere at 4 bar, rotated with a speed of 15/min. From the single crystal slab, a 1 mm thick slice in the [111] orientation was sawn and high quality surface finished [CRYSTEC, Düren, Germany].

Part of the single crystal was crushed into powder and then analyzed with x-ray diffraction. The diffractometer was a Philips Xpert equipped with a heating chamber and a platinum foil that served as a heating stage. The platinum Bragg peaks were used for the calibration of the 2θ axis. Diffractometer settings were 30 kV, 30 mA, wavelength 1.936040 Å (Fe radiation), continuous scan, 2θ step width 0.025°, 3 seconds scan time per step. The heating rate was 25 K/min. Recording time for every diffractogram was 40 minutes, during which the temperature was kept constant to ±1 K.

The DC conductivity as a function of temperature was obtained by applying the four-point method to the single crystal slab with 3 cm length using a home-built furnace with computer controlled atmosphere and temperature.

The single crystal disk with [111] orientation was characterized with x-ray photoemission spectroscopy (PES) with 450 meV excitation energy at beamline 9.3.1 at the Advanced Light Source in Berkeley, California. Prior to the PES, the sample was subject to argon ion bombardment for 30 minutes at 500 eV, a subsequent heating cycle to 500 K and back to ambient temperature, and a scan for residual carbon on the surface. This procedure was repeated two times, after which no carbon signal was detected on the crystal surface. Survey scans were made from -10 to +150 eV binding energy at temperatures from 323 K to 723 K in steps of 50 K.

X-ray absorption spectra (XAS) of the Fe K edge (ca. 7112 eV) were recorded on bending magnet beamline BM29 of the European Synchrotron Radiation Facility (ESRF) in Grenoble, France. For the transmission measurements with in situ heating in air, a pressed disk of LSF10 diluted in BN was held in a graphite envelope that was heated resistively from room temperature to 423, 573, 673, and 773 K. The feed-back thermo-couple was in contact with the sample disk and an iron foil was measured simultaneously for accurate energy calibration.

**Results**

The conductivity for LSF10 between 300 K and 1173 K is displayed in Figure 1. Increasing conductivity is observed for the temperature range from 300 K to 700 K, which follows the typical path of thermally activated small polaron hopping, $\sigma \cdot T \sim \exp(-E_P/kT)$, with activation energy of $E_p$=0.317 eV. The hopping energy for small polaron oxides is $W_H \sim E_p - E_\alpha$, where $E_\alpha$ is potential difference of lattice distortions with, and without hole. LSF30, for instance, has an activation energy of $E_p$=302 meV in a comparable temperature range [8]. The Figure shows the conductivities obtained from the single crystal as well as from a slab of pressed, sintered powder of the same material. Data obtained from single crystals may be anisotropic in nature, whereas data from sintered material may have resistivity contributions from grain boundaries. The conductivity of the single crystal attains its maximum at about 700 K, but we notice an increase of the scattering of the conductivity data points where the conductivity fluctuates between 78 S/cm and 121 S/cm. The sintered slab has a smaller conductivity maximum (82 S/cm – 93 S/cm), located at a lower temperature, 635 K. The conductivity variation with temperature shows a slight hysteresis, which is due to oxygen loss at elevated temperatures and diffusion limited re-equilibration during the cooling phase.

LSF10 adopts under ambient conditions the orthorhombic structure and undergoes a phase transformation to rhombohedral symmetry at about 700 K [1,2,6]. Figure 2 shows the X-ray powder diffraction pattern of the powderized single crystal obtained at temperatures from 313 K to 773 K. The structural changes with increasing temperature can be monitored by the splitting of Bragg reflections. A slight shoulder evolves at about 573 K at 2θ = 41.35°, which becomes more pronounced with increasing temperature. At 773 K the (020) reflex is split to equal amplitude, and separated by 0.2°. The general trend for the diffractogram is that the diffracted intensity shifts towards larger angles, indicating a decrease of lattice spacings and unit cell volume.

Figure 3 shows the Fe K edge absorption spectra between 310 K and 773 K. We see noticable changes for the three spectra at 310 K, 423 K and 573 K. However, the spectra at 573 K and 673 K differ insofar as the 673 K spectrum has a smaller chemical shift and hence represents a more reduced Fe species than the ones at lower temperatures. With the exception of the 773 K spectrum, the white lines all coincide at 7127 eV. The 773 K spectrum has pronounced quadrupol allowed $e_g$ and $t_{2g}$ prepeaks at 7104 and 7112 eV, indicative to $Fe^{3+}$. This conclusion is supported by the position of the absorption edges which are 7117.5 eV for 773 K and 7121 eV for 310 K. The average oxidation number of the iron in LSF10 is 3.1, brought about by 10% $Fe^{4+}$ and 90% $Fe^{3+}$, with a hole concentration of τ = 0.11. Since the chemical shift upon heating to 773 K is 3.5 eV, i.e. about 0.75 formula units, the Fe is reduced from average $Fe^{3.10}$ to $Fe^{2.35}$ [9].

Figure 4 displays the PES valence band (VB) spectra for LSF10 from 323 K to 723 K. The spectral assignment of the VB features is made the same way as exercised by Wadati et al. [10]. Upon temperature increase, we notice a decrease in the intensity of the $t_{2g}$ band (-3.5 eV) and the $e_g$ band, which is developed as a shoulder only at about -1.5 eV below the Fermi energy. However, the intensity near the Fermi energy increases slightly at $E_F < 0 < 700$ meV, i.e. intensity is redistributed towards the Fermi level. The inset in Figure 4 shows a magnification of the relevant range for the intensity shift near $E_F$. First, the spectral weight at 373 K near $E_F$ is smaller than for the lower temperature, 323 K. Second, the intensities here for the following temperatures do increase, with the exception of T = 723 K. There, the spectral intensity is just below that of 473 K, and above that of 323 K. We thus identify the spectrum at 673 K as the one with the highest spectral intensity near $E_F$. In order to quantify the intensity redistribution of the near the Fermi energy, we determined the intensity at $E_F$, and 100 meV above and 100 meV below the Fermi level, and plotted these three values for every temperature where the spectrum was recorded, see Figure 5, bottom panel. The three solid lines corresponding to the three energy positions show basically the same variation. Close inspection of the intensity variation curves reveals a correlation with some features of the conductivity behaviour, which for comparison is plotted on the top panel in Figure 5.

**Discussion**

On very close inspection of the conductivity data in the Arrhenius plot in Figure 1, we notice two peculiarities: First is a reversible small jump in the conductivity at 357 K with a small hysteresis; the conductivity drops (jumps) upon heating (cooling) by about 0.025 S/cm. This effect may possibly coincide with the valence band spectrum recorded at 373 K in Figure 4, which shows a remarkable decrease in spectral intensity near the Fermi energy. The second anomaly is a slight increase in the slope in the conductivity at about 573 K, right before the conductivity maximum, representing a slight increase of the activation energy from 317 meV to 332 meV. This second observation is reflected in the diffractogram at 573 K in Figure 2, where a clear transition from orthorhombic to rhombohedral structure is observed, suggesting that the rhombohedral structure is less conducting. In this temperature range we also notice enhanced scattering of the conductivity data points, suggesting that the material may be in a structural poorly defined transitional stage. This behaviour is not unexpected and to some extent reflected by the magnetic properties in this temperature range. We recollect here that LSF is predominantly antiferromagnetic, but may have canted ferromagnetic structure. We do not have any high temperature magnetic data of our sample, but our single crystal shows ferromagnetic behaviour at ambient temperature and at 373 K, as evidenced by almost rectangular magnetization curves. The Néel temperature for LSF10 is around 600 K [1,11]. Out of the LSF series, LSF10 has the highest internal magnetic field at ambient temperature [11], and the ferromagnetic Curie point may likely be at a temperature lower than 600 K. Summarizing for T ≈ 573 K, we believe that the structural transition increases the lattice distortion, impairing the Fe(3d)-O(2p) hybridization via the super-exchange angle $\theta$, and thus increasing the band width $W \sim cos^2\theta$. Consequently the band gap increases [4]. We speculate that at this tempera-ture the ferromagnetic Curie point is already passed, and that no double exchange related conductivity decrease takes place.

The conjugated intensity increase in the photoemission spectra near the Fermi energy with increasing temperature shows two extra features, this is the smaller spectral weight at 373 K than at 323 K, and the subsequent increasing spectral weight, with a subsequent maximum at about 700 K, where the conductivity has also the maximum. The back switching spectral weight at this temperature is hence paralleled by the maximum in the conductivity data.

For the quest of the physical origin for the conductivity decrease above 700 K, the temperature dependant x-ray absorption near-edge structure (XANES) spectra in Figure 3 may be helpful. The formal Fe valence in LSF10 is 3.1, about 10% $Fe^{4+}$ and 90% $Fe^{3+}$. The $Fe^{4+}$ provides the holes for the electronic transport. The XANES spectra thus resemble to a large extent the $Fe^{3+}$ features, as evidenced by the pre-edge peaks of $Fe^{3+}$ XANES spectra.

To the best of our knowledge, our work is the first higher temperature application of valence band spectroscopy on this kind of material, but there exist some key works on the use of this technique for the study of electronic properties of LSF with respect to transport properties at ambient and lower temperatures. Matsuno et al. [12] found a decrease of spectral VB PES weight near $E_F$ towards x=0.67, the well known charge ordering stage of LSF. $SrFeO_3$ has the highest spectral weight at $E_F$, compared to those by LSF. Their temperature dependent studies [12] from 130 K to 210 K indicate that the spectral changes are most pronounced for the charge ordered composition *x*=0.67 at the charge ordering transition temperature of 190 K. Wadati et al. [13] studied LSF thin films grown in-situ with VB PES and O(1s) X-ray absorption spectroscopy (XAS). They identify in the VB PES three main structures as from Fe(3d)-O(2p) bonding states, and $t_{2g}$ and $e_g$ states from Fe, the latter of which is next to the Fermi energy $E_F$. In the VB spectra recorded at room temperature, the structure nearest to $E_F$, the $e_g$ band, becomes weaker and moves toward $E_F$ as *x* is increased. In a temperature dependant study of thin films (10 K – 260 K), Wadati et al. [14] find that hole doping induces spectral weight transfer from below $E_F$ to above $E_F$ across the band gap in a highly non-rigid-like-band-manner. With increasing temperature, the spectral weight of the $e_g$ band, located about -1.5 eV from $E_F$, decreased noticeably, and there was a conjugated increase of spectral weight at about -0.2 eV from $E_F$ [10].

The three parallel dotted lines in the bottom part of Figure 5 reflect the global trend of the increase of the spectral weight around the Fermi level with increasing temperature. The spectral weight maximum at about 700 K conincides with the conductivity maximum, and the two spectral weight minima conincide with the and the change of activation energy and the jump of conductivity at about 573 K and 357 K. The intermediate *maximum* of the spectral weight at about 190 K is insofar not a maximum that would correspond to a particular high conductivity, but the continuation of the overall linear increase of the spectral weight along with the temperature increase. It is notewothy that the minute conductivity jump and the minute activation energy increase cause very significant changes in the trend of the spectral weight evolution during temperature change, whereas the global linear increase of the spectral weight is relatively unspectacular. The overall linear behaviour therefore accounts for the polarons as the primary response to thermal activation, whereas the distinct minima account for direct structural changes, which are secondary effects with respect to the temperature change.

**Conclusion**

The slightly strontium doped La$_{1-x}$Sr$_x$FeO$_3$ has a thermally activated conductivity up to a temperature of 673 K due to polaron hopping of holes created by the Fe$^{4+}$ ions. The conductivity increase is semi-quantitatively reflected by the valence band photoemission spectra, i.e. a spectral weight shift from the Fe e$_g$ band towards the Fermi energy. The density of states near the Fermi level is thus filled by the Fe$^{4+}$ hole states. Upon further heating at ambient oxygen partial pressure, LSF10 adopts a different, the rhombohedral symmetry, which is thermodynamically favoured at this temperature and oxygen partial pressure (0.2 atm), when some oxygen is released from the crystal lattice. Since the La and the Sr are virtually redox inert, it is the Fe oxidation state that will adjust to maintain charge balance in LSF10, i.e. the Fe$^{4+}$ will be reduced, and the hole concentration will decrease accordingly. This stage is met at about 700 K, where we notice the conductivity maximum and the subsequent conductivity decrease. The temperature dependent Fe K-edge X-ray absorption spectra display significant changes, including the chemical shift towards smaller energies, indicative of a reduction of the Fe. The corresponding valence band spectrum at 723 K shows consequently a smaller spectral weight at the Fermi energy than the other high temperature spectra. We believe that the depletion of holes due to depletion of oxygen at 700 K can be overcome by adjusting the ambient oxygen partial pressure, and stabilizing the polaron conductive orthorhombic structure at temperatures even higher than 700 K.

**Acknowledgement**

Financial support by the European Commission (MIRG # CT-2006-042095), the Swiss National Science Foundation (SNF # 200021-116688), CCMX NANCER and CCEM/OneBat is gratefully acknowledged. The ALS is supported by the Director, Office of Science, Office of Basic Energy Sciences, of the U.S. Department of Energy under Contract No. DE-AC02-05CH11231. The ESRF is acknowledged for beamtime at beamline BM29 under experiment number HE2469, with the assistance of Carmelo Prestipino.

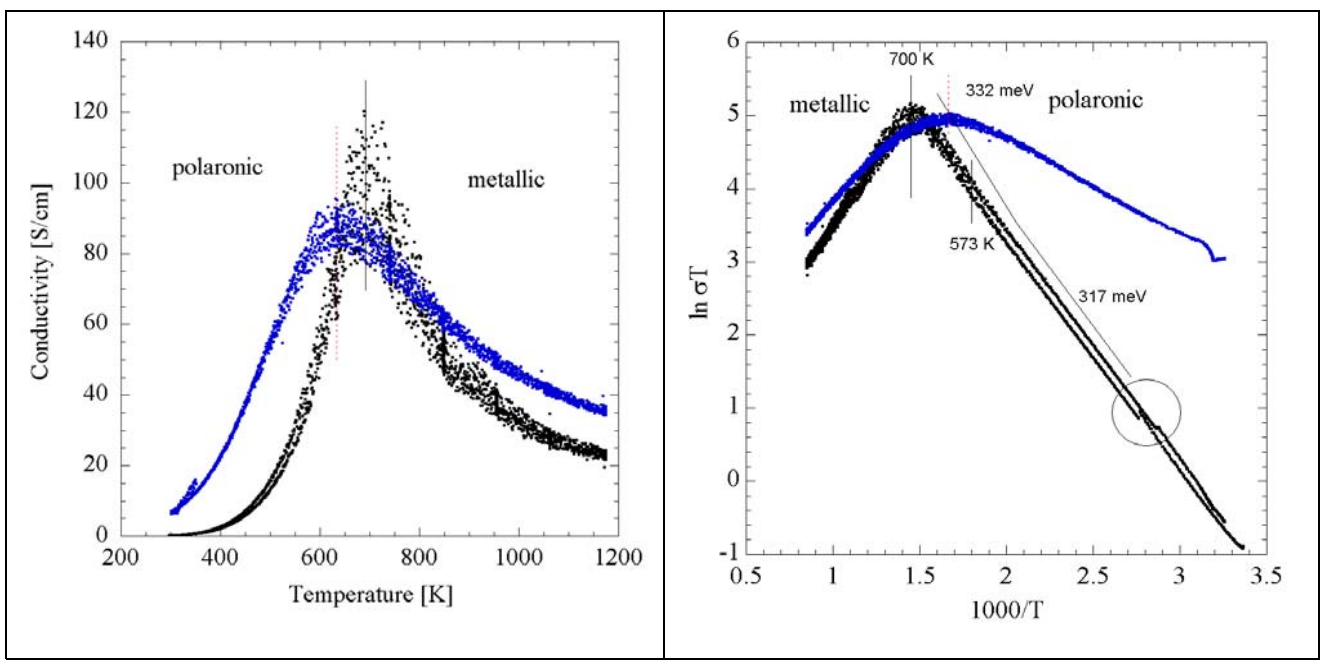

**Figure 1**: Left - Electric conductivity of the $La_{0.9}Sr_{0.1}FeO_3$ single crystal (black data points) and sintered ceramic sample (blue data points). Right – Conductivity in Arrhenius representation. Cross-over in conductivity mechanism of the single crystal at ~ 700 K.

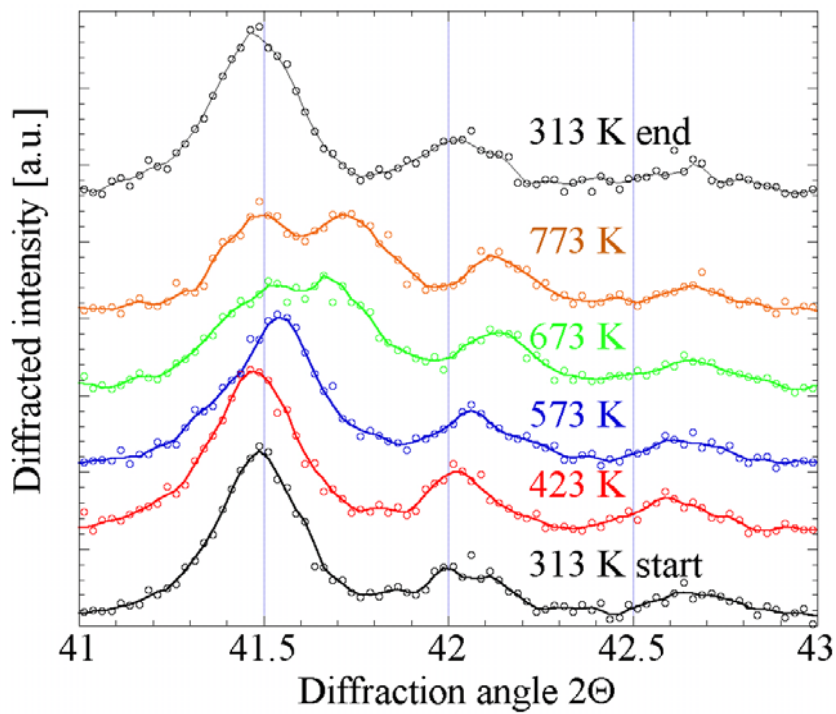

**Figure 2:** X-ray diffractogram of $La_{0.9}Sr_{0.1}FeO_3$ powder from single crystal for temperatures between 313 K and 773 K in ambient atmosphere, $\lambda$ = 1.936040 Å, Fe $K_\alpha$.

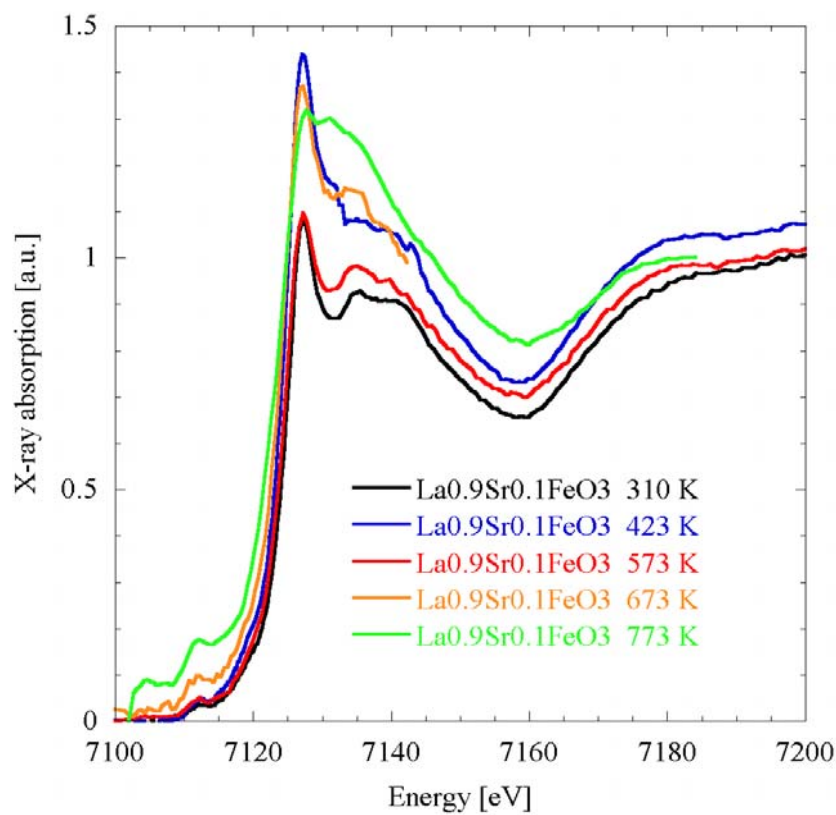

**Figure 3:** X-ray absorption near-edge structure (XANES) of LSF10 for T = 310 K – 773 K at the Fe K edge.

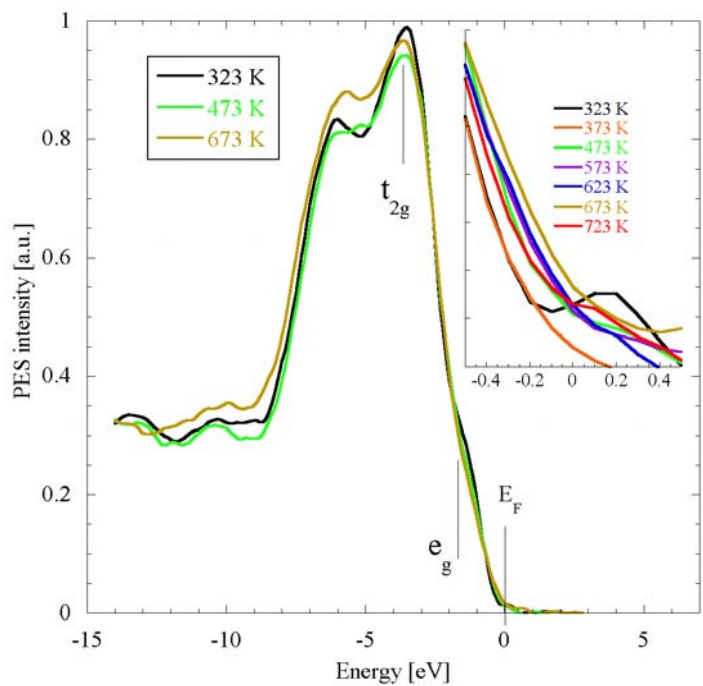

**Figure 4:** Valence band spectra of $La_{0.9}Sr_{0.1}FeO_3$ single crystal at room temperature up to 723 K (Advanced Light Source).

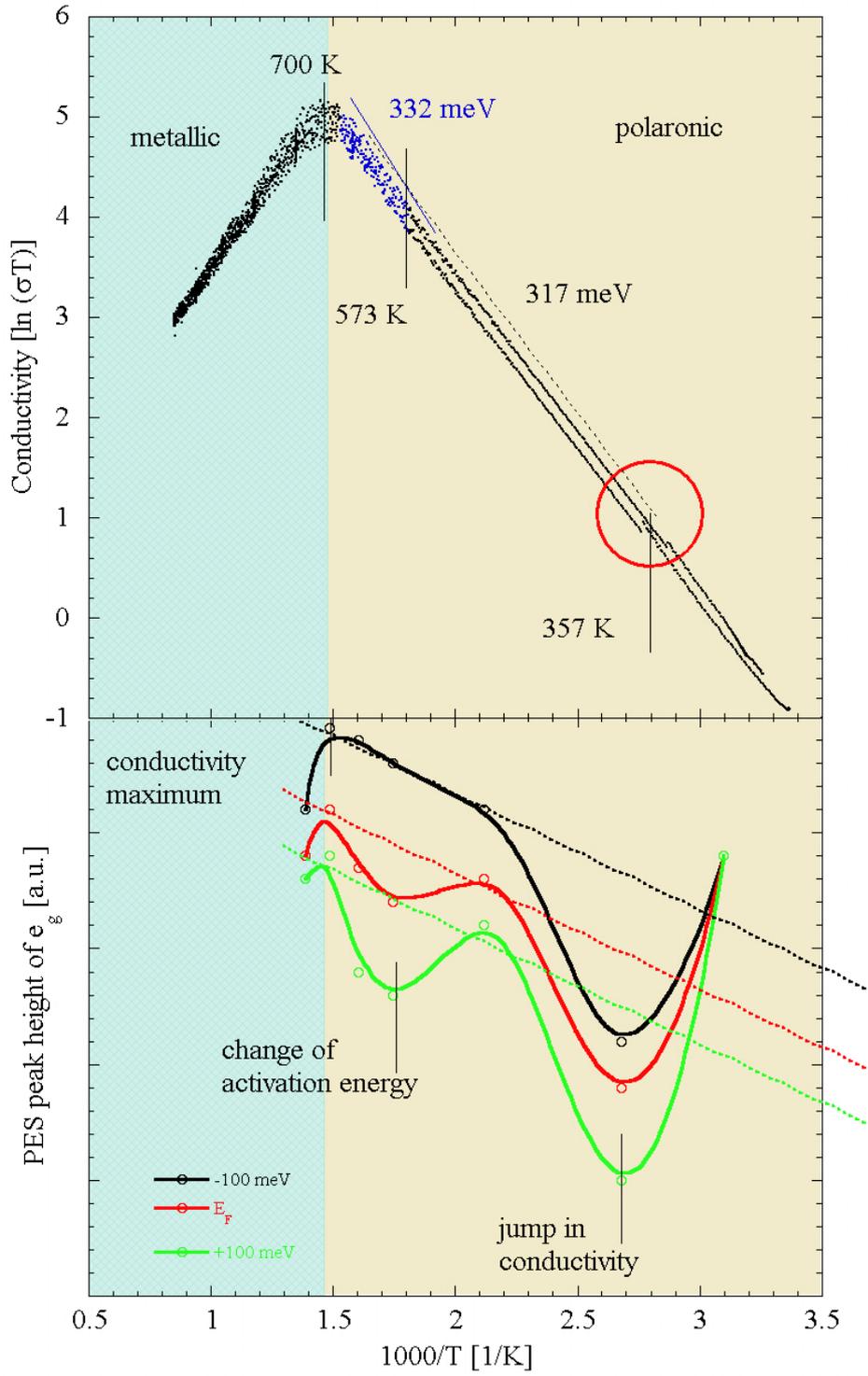

**Figure 5**: Variation of conductivity (top) and the conduction $e_g$ band height (bottom) as a function of temperature. Data are considered for $E_F$, and 100 meV above and below $E_F$.